\documentclass[epj]{svjour}

\usepackage{graphicx}

\begin{document}

\bibliographystyle{unsrt}

\title{Neutral pion production in d-Au collisions at $\sqrt{s_{NN}} = 200$~GeV}

\author{Andr\'e Mischke\thanks{\email{a.mischke@phys.uu.nl}} \it{for
the STAR Collaboration\thanks{For the full author list, see
Ref.~\cite{AziAnCorr04}.}}}
              

\institute{NIKHEF, Amsterdam, The Netherlands \\
present address: Department of Subatomic Physics, Utrecht University,
Princetonplein 5, 3584 CC Utrecht, The Netherlands.}

\date{Received: / Revised version: }

\abstract{In this paper, preliminary results are presented on high
$p_{\rm T}$ inclusive neutral pion measurements in d-Au collisions at
$\sqrt{s_{\rm NN}} = 200$~GeV in the pseudo-rapidity range $0<\eta<1$.
Photons from the decay $\pi^0\rightarrow\gamma\gamma$ are detected in
the Barrel Electromagnetic Calorimeter of the STAR experiment at RHIC.
The analysis of this first BEMC hadron measurement is described in
detail. The results are compared to earlier RHIC findings.
Furthermore, the obtained $\pi^{0}$ invariant differential cross
sections show good agreement with next-to-leading order (NLO)
perturbative QCD calculations.}

\PACS{25.75.--q}

\maketitle

\section{Introduction}
\label{intro}

After five years of data taking, exciting new observations have been
made at the Relativistic Heavy Ion Collider at Brookhaven National
Laboratory.  The high center-of-mass energy ($\sqrt{s_{NN}}$~=
200~GeV) opens up the hard scattering regime which is accessed by the
measurements of particle production at large transverse momentum.
These particles originate from parton fragmentation in the early stage
of the collisions. Hence, the scattered partons can be used to probe
the produced medium of strongly interacting matter.

A significant suppression of high $p_{\rm T}$ hadron production
relative to a simple binary collisions scaling from proton-proton has
been observed in central Au-Au collisions at RHIC~\cite{SuppPart03}.
Additionally, it was found that jet-like correlations opposite to
trigger jets are suppressed and that the azimuthal anisotropy in
hadron emission persists out to very high transverse
momenta~\cite{BBCorr03,Flow03,AziAnCorr04}.  In contrast, no
suppression effects were seen in d-Au collisions~\cite{dAuStar03},
which provide an important control measurement for the effects in cold
nuclear matter.  The d-Au measurements have led to the conclusion that
the observations made in Au-Au are due to the high density medium
produced in such collisions.  The most probable explanation to date is
parton energy loss from induced gluon radiation (jet quenching) in the
extremely dense medium. For a recent review see Ref.~\cite{Jac04}.

To quantitatively understand the existing modification from cold
nuclear matter, precise measurements of identified hadrons at high
$p_{\rm T}$ in d-Au are required.  The STAR Barrel Electromagnetic
Calorimeter (BEMC) allows high transverse momentum measurements of
$\pi^0$, $\eta$ mesons and direct photons and may contribute to the
identification of $\rho$ mesons. We present preliminary measurements
of neutral pion production in d-Au collisions.

\section{The STAR Electromagnetic Calorimeter}
\label{sec:2}

The STAR detector \cite{NimSTAR03} is one of the two large-scale
experiments at RHIC. The detection of photons is performed using the
BEMC~\cite{NimEmc03}.  This detector is one of the current upgrades of
the STAR experiment. For the RHIC Run III, 50\% of the BEMC was
installed and operational, covering an acceptance of $0<\eta<1$
(deuteron direction) and full azimuth. At the end of 2004, 35\% of the
other half of the detector ($-1<\eta<0$) 
was installed. The last stage of installation will be finished in
2005.  The STAR detector utilizes the BEMC as a leading particle
trigger to study high $p_{\rm T}$ hadron production (e.g.\ $\pi^0$ and
$\eta$) as well as rare probes (jets, direct photons and heavy quarks)
over a large acceptance range.

The BEMC is a lead-scintillator sampling calorimeter with a depth of
21 radiation lengths and an inner radius of 220~cm.  The calorimeter
covers a total area of 60~m$^2$ and is divided into 120 modules. Each
one of the EMC modules consists of 40 towers of granularity
$(\Delta\eta, \Delta\phi) = (0.05, 0.05)$.  Two layers of gaseous
shower maximum detectors (SMD), located approximately at a depth of 5
$X_0$ inside the calorimeter module, measure the lateral extension of
an electromagnetic shower and the position with high resolution
$(\Delta\eta, \Delta\phi) = (0.007, 0.007)$.  The hadron/$\gamma$
discrimination (e.g.\ $\pi^0/\gamma$) is significantly improved by
measuring the longitudinal shower shape of the tower cluster with a
pre-shower detector (PSD) located within the first two layers. The
dynamic range of the BEMC for photon detection is approximately 1 --
25~GeV/$c$ taking into account the discrimination of high transverse
momentum photons from merging decay photons.

Beam test results~\cite{NimAbsEnCal02} provide the absolute energy
calibration, whereas the relative calibration is obtained from the
peak position of minimum ionizing particles (mostly charged hadrons)
on a tower-by-tower basis~\cite{EtSTAR04}.  Moreover, an overall gain
calibration was obtained by comparing the momentum of electrons
identified in the TPC with the energy deposited in the BEMC.  The
intrinsic energy resolution due to sampling fluctuations is $\delta
E/E \approx 16\%/\sqrt{E}$.

\subsection{Event trigger}
\label{sec:21}

In the minimum bias trigger a neutron signal was required in the Zero
Degree Calorimeter (ZDC) in the Au beam direction resulting in an
acceptance of $(95\pm3)\%$ of the \mbox{d-Au} hadronic cross
section. To enhance the high $p_{\rm T}$ range, two high tower
triggers (HT1 and HT2) were used with an energy threshold of 2.5 GeV
and 5 GeV for the highest EMC cluster energy, respectively.  The tower
occupancy is in the order of a few percent for d-Au events, and the
high tower trigger efficiency is nearly 100\%.

\section{Neutral pion analysis}
\label{sec:3}

As mentioned in section 2, the installed component of the BEMC was
used for this analysis, resulting in a coverage of $0<\eta<1$ and full
azimuth ($\Delta\phi=2\pi$).  The STAR Time Projection Chamber
(TPC)~\cite{NimTPC03}, which is situated in front of the BEMC, has
full azimuthal coverage and an acceptance of $|\eta|<1.2$ for
collisions in the STAR detector centre. The TPC provides particle
identification and precise measurement of the charged particle
trajectories in a 0.5~Tesla solenoidal magnetic field.  In this
analysis, it is used as a charged particle veto.

For the analysis presented in this paper, 10\,M d-Au events 
were used after event quality cuts (main vertex coordinate (beam axis)
within 80~cm of the TPC centre). Neutral pions were reconstructed in
the decay channel $\pi^0\rightarrow\gamma\gamma$ (branching fraction
98.8\%) by calculating the invariant mass of all pairs coming from
neutral clusters in the calorimeter which do not have a TPC track
pointing at them. Furthermore, a cut on the two-particle energy 
asymmetry \mbox{$|E_1-E_2|/(E_1+E_2)$} $\le$ 0.5 was imposed.

At present, the full calibration of the calorimeter and the search for
noisy and dead towers is in progress. To perform a neutral pion
analysis at this stage, a sub-sample of good towers was selected.  The
quality of the individual towers was checked using the $\pi^0$
invariant mass distribution. The towers are tagged according 
to the decay photon with the highest energy. While determining the
gain for a given tower from $\pi^0$ peaks it is assumed that the error
from the yet uncalibrated towers used as partners for the invariant
mass analysis cancel on average.
Only those towers were used which have a well defined $\pi^0$ signal
above the combinatorial background. A Gaussian fit to the $\pi^0$
signal has to have a relative mass and width error of less than 30\%
and 50\%, respectively. By this method, approximately one third of all
towers was used for the present analysis. The acceptance correction
takes this into account. The $\pi^0$ peak position of all good towers
was used to perform an additional tower gain correction (7\% on
average).

\begin{figure}[ht]
\centering
\resizebox{0.48\textwidth}{!}{%
\includegraphics*{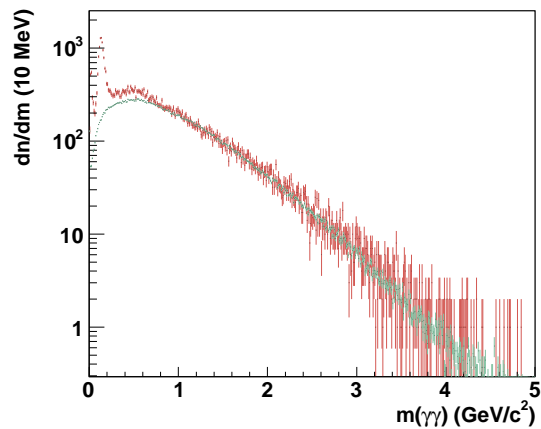}}
\resizebox{0.3\textwidth}{!}{%
\includegraphics*{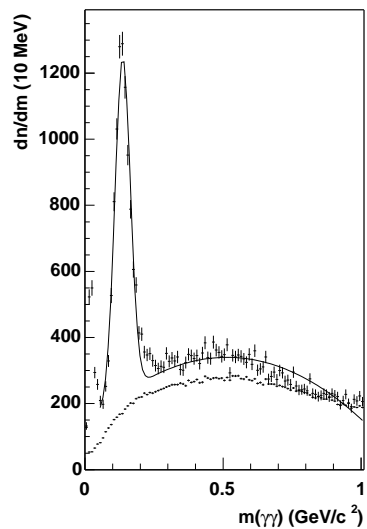}}
\caption{Invariant mass distribution of neutral pion candidates in
minimum bias d-Au collisions at $\sqrt{s_{\rm NN}} = 200$~GeV (top).
A detailed view of the range 0--1~GeV/$c^2$ is shown on the bottom
panel.  The full line is a Gaussian plus second order polynomial fit.
The histogram is obtained from event mixing.
The relative mass resolution is 20\%.}
\label{fig:1}
\end{figure}

In Fig.~\ref{fig:1}, the resulting invariant mass distribution of
neutral cluster pairs is shown. A clear signal is observed with a RMS
width of 28 MeV/$c^2$.  The background from random pairs was estimated
by two different methods. First, a second-order polynomial was fitted
to the invariant mass distribution outside the peak region (full curve
in the bottom panel of Figure~\ref{fig:1}).  With the second method,
the background is estimated using the event mixing technique.  The
event mixing distribution, which is shown in the lower histogram in
the plots of Fig.~\ref{fig:1}, follows the data in the range
0.8--5~GeV/$c^2$.  At lower invariant mass the excess can be
attributed to the tail of the $\pi^0$ peak and a contribution from the
$\eta$ signal ($m_{\eta} = 547.3$~MeV/$c^2$). The peak observed at 
$m < 0.05$~GeV/$c^2$ stems from cluster splittings in the EMC towers.

The yields per event obtained from both subtraction methods were
extracted in $p_{\rm T}$ bins (width of 0.5 GeV/$c$ for minimum bias
and 1~GeV/$c$ for high tower triggered events) by integrating the
background subtracted mass distribution in a range $\pm\,3\,\sigma$
around the $\pi^0$ peak.  The mean values were used for further
analysis and the difference contributes 10--15\% to the systematic
uncertainties. Further background studies will decrease this
systematic error contribution in the future.  The signal-to-background
ratio increases from 1 to 8 between $p_{\rm T}=1$ and 4~GeV/$c$.

Corrections for reconstruction losses (quality cuts) and detector
efficiencies were calculated with Monte-Carlo simulations using the
STAR detector geometry and reconstruction software.  A correction for
the unmeasured trigger fraction, which is expected to be a few
percent, is not applied.  Losses due to cluster density effects and
contributions from weak decays of K$^0$ mesons are not corrected for,
but are expected to be small.  The high tower trigger spectra are
normalized using pre-scaling factors obtained from the ratios of the
BEMC cluster transverse energy distributions in the overlap region.
The overall systematic errors related to efficiency, yield extraction,
pre-scaling factors, and energy calibration are estimated to be 30\%
(50\%) for transverse momenta below (above) 9 GeV/$c$.

\subsection{Results and Conclusion}
\label{sec:4}

\begin{figure}[ht]
\centering
\resizebox{0.48\textwidth}{!}{%
\includegraphics*{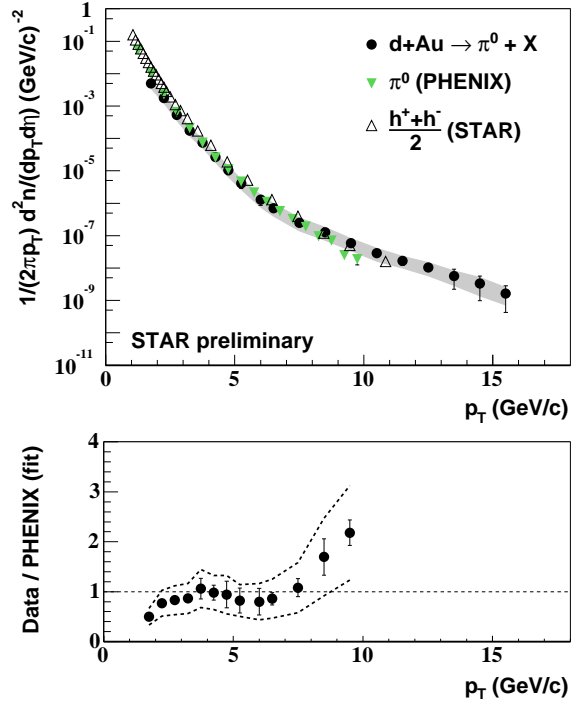}}
\caption{Inclusive $p_{\rm T}$ distribution for neutral pions in d-Au
collisions at $0<\eta<1$ (full circles). The error bars (shaded band)
represent the statistical (systematic) uncertainties.  Previous STAR
measurements of the charged hadron cross section are shown by the open
triangles, and the full triangles show $\pi^0$ results from PHENIX.
The lower plot shows the ratio between the d-Au data and a fit of the
PHENIX results, with the dashed lines indicating the systematic errors
on the data.}
\label{fig:2}
\end{figure}

The inclusive $p_{\rm T}$ distribution for neutral pions is plotted in
Fig.~\ref{fig:2}.
The yields up to 6~GeV/$c$ are from minimum bias events while above 6
(9.5)~GeV/$c$ the entries are from HT1 (HT2) triggered events. For the
different trigger samples, the yields have an overlap of one point in
the $p_{\rm T}$ spectrum and agree within errors.  It is seen from
Fig.~\ref{fig:2} that $\pi^0$ mesons are presently measured up to
$p_{\rm T}\approx16$~GeV/$c$. The obtained $p_{\rm T}$ spectrum is
compared with previous STAR measurements on charged hadron cross
sections~\cite{dAuStar03} and with PHENIX results on neutral pion
production~\cite{dAuPhenix03}.  Except for the lowest $p_{\rm T}$
point, the data show reasonable agreement within 10--20\%.

\begin{figure}[ht]
\centering
\resizebox{0.48\textwidth}{!}{%
\includegraphics*{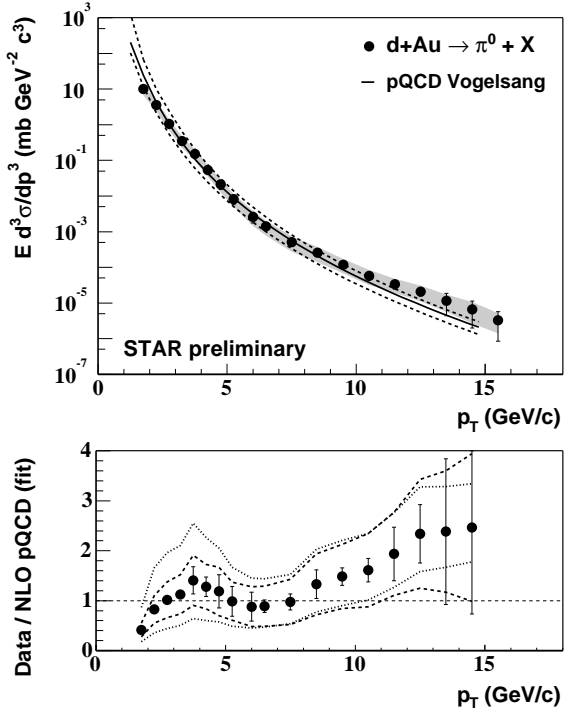}}
\caption{The invariant differential cross section (full circles)
compared to NLO pQCD calculations (full line).  The factorization
scale uncertainty is indicated by dashed lines.  The lower plot shows
the ratio between the data and the theory curve.  The dashed lines
indicate the systematic errors on the data while the dotted ones
represent different factorization scales.}
\label{fig:3}
\end{figure}

Figure~\ref{fig:3} shows the invariant differential cross section
which is obtained by multiplication of the measured yields with the
hadronic cross section in d-Au collisions ($\sigma_{\rm hadr}^{\rm
dAu} = 2.21\pm0.09$~b)~\cite{dAuStar03}.  The normalization
uncertainty is 10\%.  The results are compared to next-to-leading
order (NLO) pQCD calculations~\cite{Vog04} done using the CTEQ6M set
of nucleon parton distribution functions~\cite{CTEQ6M} and the nuclear
parton densities in the gold nucleus from Ref.~\cite{AuPDF}.  In this
calculation, the factorization scale was identified with $p_{\rm T}$
(full line) and is varied by a factor two to estimate the scale
uncertainties (dashed lines). The fragmentation functions are taken
from Ref.~\cite{KKP00}.  The Cronin effect is not included in the
calculations.  The measurements are, within errors, consistent with
the calculation up to $p_{\rm T} = 15$~GeV/$c$.

\section*{Acknowledgements}

We thank the RHIC Operations Group and RCF at BNL, and the NERSC
Center at LBNL for their support. This work was supported in part by
the HENP Divisions of the Office of Science of the U.S.  DOE; the
U.S. NSF; the BMBF of Germany; IN2P3, RA, RPL, and EMN of France;
EPSRC of the United Kingdom; FAPESP of Brazil; the Russian Ministry of
Science and Technology; the Ministry of Education and the NNSFC of
China; Grant Agency of the Czech Republic, FOM and UU of the
Netherlands, DAE, DST, and CSIR of the Government of India; Swiss NSF;
and the Polish State Committee for Scientific Research.



\end{document}